\documentclass[preprint,showpacs,aps]{revtex4}
\begin{document}
\tolerance=5000
\def\be{\begin{equation}}
\def\ee{\end{equation}}
\def\bea{\begin{eqnarray}}
\def\eea{\end{eqnarray}}

\title{On the non-existence of totally localised intersections of D3/D5 branes in
 type IIB SUGRA}
\author{Leonardo Pati\~no}
\email{e.l.patino-jaidar@durham.ac.uk}
\author{Douglas Smith}
\email{douglas.smith@durham.ac.uk}
\affiliation
   {Department of Mathematical Sciences \\
    University of Durham  \\
    Durham, DH1 3LE, U.K. }
\date{\today}

\begin{abstract}
In the present paper we study the most general configuration of intersecting D3/D5 
branes in type IIB supergravity satisfying Poincar\'e invariance in the directions common 
to the branes and SO(3) symmetry in the totally perpendicular directions. The form of
 these configurations is greatly restricted by the Killing spinor equations and the 
equations of motion, which among other things, force the Ramond-Ramond scalar to be 
zero and do not permit the existence of totally localised intersections of this kind.
\end{abstract}

\maketitle

\section{Introduction}
The configurations of intersecting branes have proved to be of great importance in 
several applications in string theory and supergravity \cite{DouglasIB}. A very well 
known example of this is the Hanany-Witten construction \cite{Hanany-Witten} which 
describs a large class of supersymmetric gauge theories, where features such as the 
running of the gauge coupling are explained in terms of geometrical aspects of the 
branes configuration.

It would therefore be desirable to enhance our present understanding of intersecting 
branes solutions in supergravity, and the present paper is an effort in this direction, 
since the existence or non-existence of totally localised solutions plays an important 
role in this understanding.

Intersections of D3/D5 branes in type IIB supergravity have been studied in the past 
\cite{Ansar}. In that occasion to approach the problem, the Ramond-Ramond scalar was assumed 
to vanish as well as some of the components of the three and five-forms field strengths 
sourced by the branes. Partial results were found pointing towards the non-existence 
of fully localised branes intersections of this kind.

In the present work, other than Poincar\'e invariance on the directions commune to the branes 
and SO(3) symmetry in the totally perpendicular directions, we don't have any initial 
assumptions about the solutions. So all the remaining restrictions, including the 
vanishing of the Ramond-Ramond scalar, come directly and exclusively from the Killing 
spinor equations and the equations of motion for the form fields involved in the 
configuration.
As a consequence we find a slightly modify set of equations which imply the non existence 
of totally localised intersections of this kind.

\section{Brane probes}

In the brane probe approximation the back reaction of the bulk to the presence of a brane 
is neglected \cite{DouglasIB}. Henceforth in this approximation, for a given 
supergravity solution, the introduction of a brane of the kind which is sourcing it 
shouldn't break further supersymmetry and in particular worldvolume supersymmetry should 
be preserved. For this to be the case the number of on-shell fermionic and bosonic 
degrees of freedom have to match and this requires the worldvolume action to be 
$\kappa$-symmetric. In the superembedding approach the imposition of this symmetry projects 
out half of the fields associated with the fermionic coordinates of the embedding of the 
world volume in the bulk.

In particular for a p-brane extending in the directions $01...p$ this projection condition reads 
\cite{Polchinski1, Polchinski2}
\be
\widehat{\Gamma}_{01...p}\epsilon_R=\epsilon_L, \label{genpc}
\ee
where $\epsilon_{R,L}$ are chiral spinors standing for the supersymmetry variation 
parameters and in general in this paper, hatted objects will represent flat space elements.

The condition (\ref{genpc}) has to hold for the supergravity solution sourced by the type 
and orientation of brane associated with it, and in the next section we'll see how to use 
this to find the supergravity solutions without the need of solving Einsteins equations.

\section{The variation of the fermionic fields}

An important property of branes and intersecting branes configurations is that they 
preserve a fraction of the supersymmetry. The precise fraction has to be found for each
case and we will do this for ours in the next section, but we will describe the 
way it will be useful to us in the following paragraphs along the lines of \cite{Douglas&Ansar}.

In a bosonic background, the variation of the bosonic fields vanishes, therefore 
to explore the amount of supersymmetry preserved by the solution we have to consider 
only the variation of the fermionic fields.
For type IIB supergravity we have to consider the variation of the dilatino $\lambda$ 
and the gravitino $\psi_M$ given \cite{Schwarz, Papadopolus} respectively by

\be
\delta\lambda=\frac{i}{\kappa}\Gamma^MP_M\epsilon^*-\frac{i}{24}
\Gamma^{MNP}G_{MNP}\epsilon, \label{vardil}
\ee
\be
\delta\psi_M={\cal{D}}_M\epsilon\equiv D_M\epsilon + U_M\epsilon + V_M\epsilon^*,\label{vargrav}
\ee
with
\bea
D_M&=&(\partial_M + \frac{1}{4}\omega^{\,\,\,\,\,\,ab}_M\widehat{\Gamma}_{ab}-
\frac{i}{2}Q_M),\nonumber\\
U_M&=&\frac{i\kappa}{48}\Gamma^{L_1...L_4}F_{ML_1...L_4}, \nonumber\\
V_M&=&\frac{\kappa}{96}(\Gamma_M^{\,\,\,\,\,\,L_1...L_3}G_{L_1...L_3}-9\Gamma^{L_1L_2}
G_{ML_1L_2}),\label{covder}\\
P_M&=&f^2\partial_MB,\nonumber\\
Q_M&=&f^2{\rm{Im}}(B\partial_MB^*),\nonumber\\
f&=&(1-BB^*)^{-1/2},\nonumber
\eea
and $B$ is related to the dilaton and the Ramond-Ramond scalar via the equation
\be
C_0+i\tau_2=i(\frac{1-B}{1+B}). \label{rrscB}
\ee

Because of its convenience for the rest of the paper,  the chiral spinors 
$\epsilon_{R,L}$ parameterising the variation have been written as a single complex spinor 
$\epsilon=\epsilon_R+i\epsilon_L$. The connection $\omega^{\,\,\,\,\,\,ab}_M$ is the spin 
connection, and it is contracted with a {\it{flat space}} gamma matrix because $a$ and 
$b$ are flat space indexes, even though, since these are totally contracted, it is not 
necessary to write any of the other objects in flat space terms.

For the solution to be supersymmetric (\ref{vardil}) and (\ref{vargrav}) have to 
vanish. Therefore the amount of supersymmetry preserved by the solution can be obtained 
from the real dimensionality of the space spanned by the variation parameters compatible 
with the projection condition of the kind (\ref{genpc}) for the specific configuration.

A tactic to find supersymmetric configurations is to obtain the corresponding projection 
conditions of the kind (\ref{genpc}) and then impose them while solving the equations 
$\delta\lambda=0$ and $\delta\psi_M=0$, reducing therefore the number of solutions. It 
turns out often to be the case that this procedure determines the system completely, so 
even though the Einstein equations are not explicitly solved, they are automatically 
satisfied, as should be if these solutions are to exist at all.

The equations which are need to be solved in this scheme are much simpler than the 
Einstein equations, so this is how we will proceed in this paper.

\section{Intersecting D5/D3 branes}

Let's describe the system we are interested in by D5-branes extended in the 
directions $x^0,x^1,x^2,x^3,x^4,x^5$ and D3-branes along the directions $x^0,x^1,x^2,x^6$, 
so that they intersect over $x^0,x^1,x^2$ wile $x^7,x^8,x^9$ are totally perpendicular 
directions. All the branes are at the origin of the totally perpendicular 
directions, but the D5-branes can be spread over $x^6$, and the 
D3-branes over $x^3,x^4,x^5$.

For this system there is a very convenient way to write the metric \cite{Ansar}
\be
ds^2=H_1^2\,\eta_{\mu\nu}dx^\mu dx^\nu+g_{\alpha\beta}dx^\alpha dx^\beta 
+ D^2(dx^6+A_\alpha dx^\alpha)^2 + H_2^2 \delta_{ij}dx^i dx^j , \label{g}
\ee
where Greek indexes from the middle of the alphabet like $\mu$ and $\nu$ run from 0 to 2, 
Greek indexes from the beginning of the alphabet run between 3 and 5, and lower case Latin 
characters take the values 7, 8 and 9. Upper case Latin indexes will run from 0 to 9.

Notice that on writing this metric we are assuming SO(3) symmetry on the totally perpendicular 
directions. If we take the metric to be independent of the common 
directions, then we will also have 2+1D Poincar\'e invariance. Nevertheless, the 
most general metric on the partially perpendicular directions can be expressed using the second 
and third terms in (\ref{g}).

Now, the projection conditions for the D3 and D5 branes are respectively
\bea
\widehat{\Gamma}_{0126}\epsilon_R&=&\epsilon_L,\nonumber\\
{\mathrm{and}} \label{rpc}\\
\widehat{\Gamma}_{012345}\epsilon_R&=&\epsilon_L .\nonumber
\eea

Independently of the configuration, the spinors $\epsilon_{R,L}$ satisfy
\bea
\widehat{\Gamma}_{0123456789}\epsilon_R&=&\epsilon_R, \nonumber\\
{\mathrm{and}} \label{mw} \\
\widehat{\Gamma}_{0123456789}\epsilon_L&=&\epsilon_L, \nonumber
\eea
since they are Mayorana-Weyl fermions.

Using some Dirac algebra it is easy to see that the conditions given by (\ref{rpc}) and 
(\ref{mw}) can be succinctly written as
\bea
\widehat{\Gamma}_{0126}\epsilon&=&-i\epsilon, \nonumber\\
\widehat{\Gamma}_{3456}\epsilon&=&-\epsilon^*, \label{cpc} \\
\widehat{\Gamma}_{7896}\epsilon&=&-i\epsilon^* ,\nonumber
\eea
in terms of the complex spinor $\epsilon$ introduce earlier. This way of writing the 
projection conditions proved to be very convenient.

This is a good point to remember that we are not assuming anything about the five-form and 
three-form field strengths, so any restriction to their components will come from the 
corresponding equations of motion and the Killing spinor equations.

\section{The Killing spinor equations}

What follows is to write down explicitly the equations $\delta\lambda=0$ and 
$\delta\psi_M=0$, then by repeated use of (\ref{cpc}) and Dirac manipulations get to 
extract the coefficients of the linearly independent combinations of $\widehat{\Gamma}$ 
matrices acting on $\epsilon$ and independently those acting on its complex conjugate 
$\epsilon^*$ \cite{DouglasIB}.

A lengthly and not particularly enlightening calculation reveals the following. 
We can start with $\delta\psi_M=0$ and pick the coefficients of the two indexes 
gamma matrices $\widehat\Gamma_{MN}$. It is worth noticing that these coefficients involve 
neither the dilaton nor the Ramond-Ramond scalar at all. From the resulting equations we see that 
the part of the metric given by $g_{\alpha\beta}$ has to be conformaly flat, and therefore we write 
it as diag$(g^2,g^2,g^2)$, $g$ being a general function.

From the same coefficients we obtain the equations
\bea
2H_1^{-1}\partial_M H_1+H_2^{-1}\partial_M H_2 + g^{-1}\partial_M g&=&0, \nonumber\\
3H_1^{-1}\partial_M H_1+2H_2^{-1}\partial_M H_2 - D^{-1}\partial_M D&=&0, \nonumber\\
4(g^{-1}\partial_\alpha g+D^{-1}\partial_\alpha D)-4(g^{-1}\partial_6 g+D^{-1}
\partial_6 D)A_\alpha -\partial_6A_\alpha&=&0, \label{kee}\\
\partial_\alpha A_\beta -\partial_\beta A_\alpha - A_\alpha \partial_6 A_\beta
+A_\beta \partial_6 A_\alpha &=&0, \nonumber\\
\partial_i A_\alpha &=&0,\nonumber
\eea
and the fact that the only non zero components of the five-form are 
$\widehat{F}_{0126\alpha}$, $\widehat{F}_{0126i}$ and those related to them by the self 
duality, $\widehat{F}_{345ij}$ and $\widehat{F}_{\alpha\beta 789}$. 
All the other components of the five-form vanish identically.

The first and second equations in (\ref{kee}) can be solved respectively by setting
\bea
g&=&H_1^{-2}H_2^{-1},  \nonumber\\
&{\mathrm{and}}&\label{solgd}\\
D&=&H_1^3H_2^2. \nonumber
\eea

Introducing $H\equiv g^4D^4$ we can recast our remaining set of equations as
\bea
\partial_\alpha H -\partial_6 (HA_\alpha)&=&0, \nonumber\\
\partial_\alpha A_\beta -\partial_\beta A_\alpha - A_\alpha \partial_6 A_\beta
+A_\beta \partial_6 A_\alpha &=&0, \label{rkee}\\
\partial_i A_\alpha &=&0. \nonumber
\eea

Using (\ref{solgd}) we see that $H=H_1^4H_2^4$, but it turns out to be useful to 
introduce $K\equiv H_1^{-2}H_2^2$, and keep $H,K$ and $A_\alpha$ as the independent 
gravitational fields.

Still from the same components of the equations, we find for the five-form
\bea
\widehat{F}_{0126\alpha}&=&\frac{\widehat{\varepsilon}_\alpha^{\,\,\,\,\beta\gamma}}{2}
\widehat{F}_{789\beta\gamma}=\frac{H^{3/8}K^{3/4}\delta_\alpha^\delta}{4\kappa}
( \partial_\delta K^{-1}-A_\delta\partial_6K^{-1}), \nonumber\\
\label{Fh}\\
\widehat{F}_{0126i}&=&\frac{-\widehat{\varepsilon}_i^{\,\,\, jk}}{2}
\widehat{F}_{345jk}=\frac{K^{1/4}\delta_i^l}{2\kappa H^{5/8}}
\partial_l(H^{1/2}K^{-1/2}) , \nonumber
\eea
and for the three form
\bea
\widehat{G}_{6\alpha\beta}&=&
\widehat{\varepsilon}_{\alpha\beta}^{\,\,\,\,\,\,\, \gamma}
\frac{\delta_\gamma^\delta}{\kappa H^{5/8}K^{1/4}}(\partial_\delta H-A_\delta\partial_6H), \nonumber\\
\widehat{G}_{345}&=&\frac{-2K^{3/4}}{\kappa H^{9/8}}\partial_6
\frac{H^{1/2}}{K^{1/2}}, \label{Gh}\\
\widehat{G}_{6ij}&=&\widehat{\varepsilon}_{ij}^{\,\,\,\,\,\, k}
\frac{-iK^{1/4}\delta_k^l}{\kappa H^{9/8}}\partial_l H, \nonumber\\
\widehat{G}_{789}&=&\frac{i}{\kappa H^{5/8}K^{3/4}}\partial_6K , \nonumber
\eea
where we have used the elements of the zehnbein in terms of $H,K$ and $A_\alpha$ to 
change from curved space-time components to flat space-time components, and the 
symbol $\widehat{\varepsilon}$ is the totally antisymmetric tensor in flat space, 
that is, we raise and lower indexes on it by using $\eta_{MN}$.

Finally from the components of the equation $\delta\lambda=0$ with one index gamma matrices 
$\widehat{\Gamma}_M$ we find
\bea
\widehat{P}_\alpha &=&\frac{-\kappa}{8}\widehat{\varepsilon}_\alpha^{\,\,\,\,\beta\gamma}
\widehat{G}_{6\beta\gamma},\nonumber\\
\widehat{P}_6 &=&\frac{\kappa}{4}(\widehat{G}_{345}+i\widehat{G}_{789}),\label{PG}\\
\widehat{P}_i&=&\frac{-i\kappa}{8}\widehat{\varepsilon}_i^{\,\,\, jk}
\widehat{G}_{6jk},\nonumber
\eea
and the result that all the remaining components of $\widehat{\mathbf{G}}$ not listed in
(\ref{Gh}) vanish.

The rest of the equations coming from the linearly independent combinations of gamma 
matrices in the Killing spinor equations are redundant.

\section{The vanishing of the Ramond-Ramond scalar}

To completely determine the system we need to solve the equations of motion for the
five-form and three-form field strengths. Let's start then by solving the equations given 
by
\be
d{\mathbf{F}}_3=0, \label{dF}
\ee
where ${\mathbf{F}}_3$ is related to ${\mathbf{G}}$ by
\be
{\mathbf{G}}=f({\mathbf{F}}_3-B{\mathbf{F}}_3^*). \label{F3G}
\ee

To this end, using (\ref{covder}), (\ref{Gh}) and (\ref{PG}) we see that
\be
f^2\partial_MB=P_M=\frac{-1}{4H}\partial_MH. \label{PH}
\ee

Since $f^2$ and $H$ are real, the imaginary part of B has to be just a constant. Writing 
$B=B_r+iB_0$ we find that the most general solution to (\ref{PH}) for constant $B_0$ is

\be
B_r=\sqrt{1-B_0^2}\frac{1-C_1H^{\frac{1}{2}\sqrt{1-B_0^2}}}
{1+C_1H^{\frac{1}{2}\sqrt{1-B_0^2}}}, \label{solBr}
\ee
with
\be
C_1=e^{\frac{C_0}{\sqrt{1-B_0^2}}}. \label{c1}
\ee

Using this expression for $B$ in (\ref{F3G}) we can recover the real and imaginary parts 
of ${\mathbf{F}}_3$,

\bea
{\mathrm{Re}}{\mathbf{F}}_3&=&\frac{1}{2\sqrt{1-B_0^2}}\left[\left(\frac{1+\sqrt{1-B_0^2}}
{\sqrt{C_1}H^{\frac{1}{4}\sqrt{1-B_0^2}}}+(1-\sqrt{1-B_0^2})\sqrt{C_1}H^{\frac{1}{4}
\sqrt{1-B_0^2}}\right)
{\mathrm{Re}}{\mathbf{G}}\right.  \nonumber \\
&+& \left. \left(\frac{1}{\sqrt{C_1}H^{\frac{1}{4}\sqrt{1-B_0^2}}}+\sqrt{C_1}H^{\frac{1}{4}
\sqrt{1-B_0^2}}\right) B_0{\mathrm{Im}}{\mathbf{G}}\right] \nonumber\\ \nonumber\\
& &{\mathrm{and}} \label{riF} \\ \nonumber\\
{\mathrm{Im}}{\mathbf{F}}_3&=&\frac{1}{2\sqrt{1-B_0^2}}\left[\left(\frac{1}{\sqrt{C_1}H^{\frac{1}{4}
\sqrt{1-B_0^2}}}+\sqrt{C_1}H^{\frac{1}{4}\sqrt{1-B_0^2}}\right)B_0{\mathrm{Re}}
{\mathbf{G}} \right. \nonumber \\
&+&\left.\left(\frac{1-\sqrt{1-B_0^2}}{\sqrt{C_1}H^{\frac{1}
{4}\sqrt{1-B_0^2}}}+(1+\sqrt{1-B_0^2})\sqrt{C_1}H^{\frac{1}{4}\sqrt{1-B_0^2}}\right)
{\mathrm{Im}}{\mathbf{G}}\right]. \nonumber
\eea

Equation (\ref{dF}) implies that the exterior derivative of these last two expressions 
has to vanish. Given the form of ${\mathbf{G}}$ it is not difficult to see that this 
vanishing can only be possible if the coefficients of the real part of $\mathbf{G}$ 
in (\ref{riF}) are identical or one of them vanishes, and the same thing applies to the
coefficients of the imaginary part of $\mathbf{G}$.
Examination of the coefficients leads to the fact that the only regular solution which 
satisfy these constrains is $B_0=0$, and since from (\ref{rrscB}) the Ramond-Ramond 
scalar is given by
\be
C_0=\frac{2B_0}{(1+B_r)^2+B_0^2}, \label{rrB}
\ee
we see that this implies the vanishing of the Ramond-Ramond scalar.

The expressions (\ref{riF}) then further simplify to
\bea
{\mathrm{Re}}{\mathbf{F}}_3=H^{-1/4}{\mathrm{Re}}{\mathbf{G}} \nonumber \\
{\mathrm{and}} \label{riFs} \\
{\mathrm{Im}}{\mathbf{F}}_3=H^{1/4}{\mathrm{Im}}{\mathbf{G}} \nonumber
\eea

We still have to solve the equations $d{\mathrm{Re}}{\mathbf{F}}_3=d{\mathrm{Im}}{\mathbf{F}}_3=0$, 
but we already see that for these solutions to exist the Ramond-Ramond scalar {\it{must}} 
vanish.

Let's finish this section by noticing that $B$ is not present in the expressions 
(\ref{riFs}) but with these further simplifications we can solve for it straight away 
from (\ref{PH}) and use (\ref{rrscB}) to write down the solution for $\tau_2$
\be
\tau_2=H^{1/2}. \label{tau2}
\ee

\section{The non-existence of totally localised solutions}

Once we have seen that the Ramond-Ramond scalar vanishes, we can write down the equations 
of motion as
\bea
d{\mathbf{F}}_5+\frac{\kappa}{4}({\mathrm{Im}}{\mathbf{F}}_3\wedge{\mathrm{Re}}
{\mathbf{F}}_3)&=&*\mbox{\boldmath $\rho$}_{D3},\nonumber \\
d{\mathbf{F}}_3&=&*\mbox{\boldmath $\rho$}_{D5}, \label{eqmo}\\
d*(\tau_2{\mathrm{Re}}{\mathbf{F}}_3)&=&4\kappa{\mathbf{F}}_5\wedge
{\mathrm{Im}}{\mathbf{F}}_3, \nonumber \\
d*(\tau_2^{-1}{\mathrm{Im}}{\mathbf{F}}_3)&=&-4\kappa{\mathbf{F}}_5\wedge
{\mathrm{Re}}{\mathbf{F}}_3, \nonumber
\eea
where $*$ is the Hodge star operator and $\mbox{\boldmath $\rho$}_{Dp}$ is the p+1-form 
given the density distribution of the Dp-branes.

To advance any further we notice that the conformal flatness we found for the part of the 
metric given by $g_{\alpha\beta}$ implies SO(3) symmetry in these directions, so if we 
introduce spherical coordinates on this three dimensional space, the only dependence on 
the coordinates will be through the radial coordinate $r$. We can do the same for 
the totally perpendicular directions, and denote the radial coordinate as $\rho$.

The result of rewriting the differential forms in terms of these coordinates is that 
after a straightforward calculation all of the equations given by the different componets of 
(\ref{eqmo}) reduce to a set of independent differential equations, being the source equations
\bea
& &\!\!\!\!\!\frac{1}{r^2}\partial_r[r^2(\partial_rK-A_r\partial_6K)]+\frac{1}{\rho^2}
\partial_\rho[\rho^2\partial_\rho(KH^{-1})]+\partial_6K[H^{-1}\partial_6(H^{-1}K)+A_r
\partial_6A_r]=\rho_{D3}, \nonumber \\
& &{\mathrm{and}} \label{surce} \\
& &\!\!\!\!\!\frac{1}{\rho^2}\partial_\rho(\rho^2\partial_\rho H)+\partial_6^2K=\rho_{D5}, \nonumber
\eea
along with the identities
\bea
(\partial_\rho H^1)\partial_6(H^{-1}K)&=&0, \label{bian1}\\
(\partial_\rho K^1)\partial_6(H^{-1}K)&=&0, \label{bian2}\\
(\partial_6A_r)(\partial_6K)&=&0, \label{bian3}\\
(A_r\partial_6K-\partial_rK)(\partial_6K)&=&0, \label{bian4}\\
\partial_\rho A_r&=&0, \label{bian5}\\
\partial_rH-\partial_6(A_rH)&=&0,\label{bian6}\\
\partial_6[H^{-1}\partial_6(H^{-1}K)+A_r\partial_6A_r]-\frac{1}{r^2}\partial_r
(r^2\partial_6A_r)&=&0,\label{bian7} 
\eea
where the $\rho_{Dp}$ are functions coming from the pertinent components of the 
p+1-forms $\mbox{\boldmath $\rho$}_{Dp}$.

It is from these equations that the non-existence of totally localised solutions is apparent, 
but let's be explicit about it.

One way to see it is to start by considering $K$ in the source equation of the D5-branes, and analyse 
the only two possible cases. 

1) The first alternative is 

\be
\partial_6K=0,
\ee
so that the distribution is purely given by $H$. In this case $\partial_\rho H\neq 0$ if we want 
D5-branes at all, and $\partial_6H\neq 0$ if these are to be localised on the $x^6$ direction. 
From (\ref{bian1}) we see that these three conditions are not compatible.

2) The second case we need to analyse is 
\be
\partial_6K\neq 0. \label{K6no0}
\ee
Localisation on $\rho$ of the D5-brane requires 
\be
\partial_\rho K\neq 0, \label{Krhono0}
\ee
which along with (\ref{bian2}) implies 
\be
\partial_6(H^{-1}K)=0. \label{HK60}
\ee

On the other hand, (\ref{K6no0}), (\ref{bian3}) and (\ref{bian4}) imply
\bea
\partial_6A_r=0, \nonumber \\
{\mathrm{and}} \label{Ar60} \\
A_r\partial_6K-\partial_rK=0. \nonumber
\eea
So we see that in the source equation for the D3-branes only the second term is different from 
zero. Localisation on $r$ of the D3-branes would require $\partial_r(KH^{-1})\neq 0$, but this is not 
the case, since equations (\ref{K6no0}), (\ref{Ar60}), (\ref{bian4}) and (\ref{bian6}) imply
\bea
\partial_rK&=& A_r\partial_6K, \nonumber \\
{\mathrm{and}} \nonumber \\
\partial_rH&=& A_r\partial_6H, \label{patialsKH} \\
{\mathrm{so}} \nonumber \\
\partial_r(KH^{-1})&=&A_r\partial_6(KH^{-1})=0. \nonumber
\eea
The last line is true because of (\ref{HK60}).

We see then that, the first case is simply inconsistent, whereas in the second the localisation 
of the D5-branes implies the smearing of the D3-branes, which completes the proof on the 
non-existence of totally localised solutions for our case.

It is very easy to see, but worth commenting, that if we set to zero one or the other brane 
distribution densities, the remaining equations lead to the known (fully localised) solutions for 
only D5 or D3 branes configurations.

\section{Why are there not localised solutions?}

The reason why no totally localised solutions exist for certain configurations of intersecting 
branes has been discussed in the past in terms of the field theory 
associated with them in the near horizon limit \cite{Don, Andres}. Unfortunately the 
configuration we are treating here has not been considered in this discussion, though it has been 
commented \cite{Don} that there could be indications of a breakdown of holography when the 
dimension of the totally perpendicular space is three, which is precisely our case. 

Notice that the configuration our computations are pointing to, is that of a totally localised 
D5-brane with a uniformly smeared D3-brane over its world volume spanning one extra dimension. 
Whether this smearing is related to a no-hair theorem for the D5-brane is far from clear to us yet, 
but this could be the case in analogy with the analysis of \cite{DonNH}.

To properly understand which is the mechanism preventing this configurations to be totally 
localised further investigation is necessary. A first steep in this direction would be to separate 
the branes along one of the totally perpendicular directions, and then analyse 
the smearing as we bring the branes closer together.

\section*{Acknowledgements}

We would like to thank Ansar Fayyazuddin for helpful comments and discussions.

\end{document}